\documentclass[12pt]{article}

\usepackage{newtxtext,newtxmath}

\usepackage{graphicx}

\usepackage[letterpaper,margin=1in]{geometry}

\renewenvironment{abstract}
	{\quotation}
	{\endquotation}

\date{February 14th, 2025}

\makeatletter
\renewcommand{\fnum@figure}{\textbf{Figure \thefigure}}
\renewcommand{\fnum@table}{\textbf{Table \thetable}}
\makeatother

\usepackage{scicite}

\usepackage{url}

\def\scititle{
	zScore: A universal decentralised reputation system for the blockchain economy
}
\title{\bfseries \boldmath \scititle}

\author{
	Himanshu~Udupi$^{1\ast\dagger}$,
	Prof.~Ashutosh~Sahoo$^{2}$,
	Akshay~SP$^{2}$,
    Gurukiran~S$^{2}$ \and
    Parag~Paul$^{3}$,
    Prof.~Petrus~C.~Martens$^{4}$ \and
	\small$^{1}$Siebel School of Computing and Data Science (UIUC)\and
	\small$^{2}$Zeru Finance\and
    \small$^{3}$Rig AI 
    \small$^{4}$ Department of Physics and Astronomy, Georgia State University \and
	\small$^\ast$Corresponding author. Email: hudupi2@illinois.edu\and
	\small$^\dagger$These authors contributed equally to this work.
}

\begin{document} 

\maketitle

\begin{abstract} \bfseries \boldmath

Modern society functions on trust \cite{misztal2013trust}. The onchain economy \cite{laul2024onchain}, however, is built on the founding principles of trustless peer-to-peer interactions \cite{nakamoto2008bitcoin} in an adversarial environment without a centralised body of trust and needs a verifiable system to quantify credibility to minimise bad economic activity \cite{laul2024onchain}. We provide a robust framework titled zScore, a core primitive for reputation derived from a wallet's onchain behaviour using state-of-the-art AI neural network models[Section 2.1,3] combined with real-world credentials ported onchain through zkTLS. The initial results tested on retroactive data from lending protocols establish a strong correlation between a good zScore and healthy borrowing and repayment behaviour, making it a robust and decentralised alibi for creditworthiness[Section 4]; we highlight significant improvements from previous attempts by protocols like Cred \cite{wolf2022scoring} showcasing its robustness. We also present a list of possible applications of our system in Section 5, thereby establishing its utility in rewarding actual value creation while filtering noise and suspicious activity and flagging malicious behaviour by bad actors[Section 5,4.4].

\end{abstract}

\section{Introduction}
\noindent
Modern society is built on the foundations of trust \cite{misztal2013trust}. For example, we trust intermediaries / middlemen whenever we make a transaction \cite{diamond1983bank}, we trust reviews on a platform while making purchase decisions \cite{chevalier2003effect}, etc. However, the onchain economy \cite{laul2024onchain}, a holistic term that refers to economic activities powered by the crypto-economic security of the blockchain infrastructure, is built in a manner facilitating peer-to-peer economic interactions in a trustless or trust-minimized fashion \cite{nakamoto2008bitcoin} to bring in efficiencies by eliminating the intermediary, often a centralised institution that acts as a body of trust between strangers \cite{diamond1983bank}, and enabling them to engage in economic activity. This fundamental difference between Decentralised Finance(DeFi) and Traditional Finance (TradFi) is the transparency of DeFi mechanisms, which acts as an alibi for trust, while TradFi mechanisms are largely proprietary and trusting intermediaries are crucial to the functioning of TradFi. \cite{jensen2021introduction}.   

As the onchain economy \cite{laul2024onchain} grows beyond its nascency, bad actors emerge, leading to economic frauds and defaults. Even as regulatory frameworks evolve globally with lawmakers devising policies to encompass interactions in the onchain economy, the need for reputation is imminent. Attaching a reputation score at a wallet level makes an additional dimension of trust available for decision-making. zScore[Section 2.1] is an effort towards building a robust, decentralised reputation layer to facilitate interactions in the onchain economy. These interactions could involve the disbursement of loans against a collateral asset, trading a cryptocurrency,  minting a digital artwork as an NFT, distributing incentives to the community by a protocol, rewarding user loyalty and much more. 

An economy must have a source of capital and a financial sphere to sustain itself \cite{buiter2014role}. For onchain economies, similar to modern society, we have Decentralized Finance (DeFi) \cite{jensen2021introduction}, which supports it. Decentralised Finance has five verticals: Lending Protocols \cite{kaplan2023decentralized}, Decentralized Exchanges \cite{lehar2021decentralized}, Perpetuals \cite{bitstamp2024perp}, Liquid Staking Tokens (LSTs) and Liquid Restaking Tokens (LRTs) [\cite{gogol2024liquid},\cite{carre2024liquid}], and OnChain Credit [\cite{3jane2024introducing},\cite{etherfi_whitepaper}]. Parallel to modern society, reputation is derived from interactions with all of these five; a person's ability to manage risk and debt drastically influences their FICO scores \cite{arya2013anatomy}; similarly, in onchain economies, a person's interaction with DeFi will influence their zScores[Section 2.1]. 

Lending protocols \cite{kaplan2023decentralized} are the largest and most influential vertical out of all the verticals in DeFi. While zScore [Section 2.1] is a complex system that quantifies reputation for lending protocols, we draw inspiration from FICO scores \cite{arya2013anatomy} while assessing the role of lending protocols \cite{kaplan2023decentralized} in determining reputation [Section 4]. In the following paragraphs we introduce lending protocols \cite{kaplan2023decentralized} and current mechanisms used to prevent defaults. 

Lending protocols in Decentralized Finance (DeFi) \cite{kaplan2023decentralized} share similarities with Traditional Finance (TradFi) lending architectures; users in both sectors can borrow assets by pledging some percentage of their assets as collateral. However, lending protocols do not have atomic loans \cite{wolf2022scoring}, making it difficult to calculate risk profiles; we tackle this using our definition of zScore [Section 2.1] in Section 4. The key difference, however, is the absence of a credit scoring model in DeFi lending, which has resulted in inefficient capital allocation for both the lender and the borrower, as discussed in the following paragraphs. 

Lending protocols in DeFi often combat the absence of a measure of risk, like a credit scoring model, by requiring users to pledge approximately 150\% of the loan value as collateral \cite{rareskills_defi_liquidations}. This system is a deterrent to default but has also resulted in US\$15 billion worth of assets lying idle \cite{unsecured_business_loans_market_report_2024} Another method lending protocols employ to protect lenders is setting low loan-to-value ratios (LTV), resulting in a lot of idle capital, as discussed in \cite{unsecured_business_loans_market_report_2024}. Additionally, the unsecured credit market is valued at US\$4.5 trillion globally and is projected to grow with a compounded annual growth rate of 11.3\% to a US\$7.67 trillion economy by 2028 \cite{unsecured_business_loans_market_report_2024}. The absence of a credit scoring mechanism in DeFi will limit the growth of the onchain economy, restricting users to traditional finance lending architectures, which are not decentralised and have much room for bias \cite{ioannou2021financial}. Capital inefficiency highlights the need for a native credit-scoring solution built for DeFi. This paper presents our attempt at building and applying a credit scoring model based solely on Aave V3 protocol \cite{frangella2022aave} usage. There have been several attempts before this by others–Cred \cite{wolf2022scoring}, Rocify \cite{rocifi_documentation}, Chainrisk \cite{ghosh2024chain}, etc.–and we build off their findings to introduce our methodology, which, we argue, is more robust and adaptive than previous approaches. 

\textbf{Our objectives} of this paper are to: (i) Provide a novel, robust model which quantifies onchain reputation using zScores[Section 2.1], (ii) Apply our model to a singular DeFi vertical of lending protocols and demonstrate the model's utility and advantages [Section 4], (iii) Provide multiple incentive schemes which utilise zScores to improve capital efficiency across the onchain economy. 

\begin{figure}[ht] 
	\centering
	\includegraphics[width=0.7\textwidth]{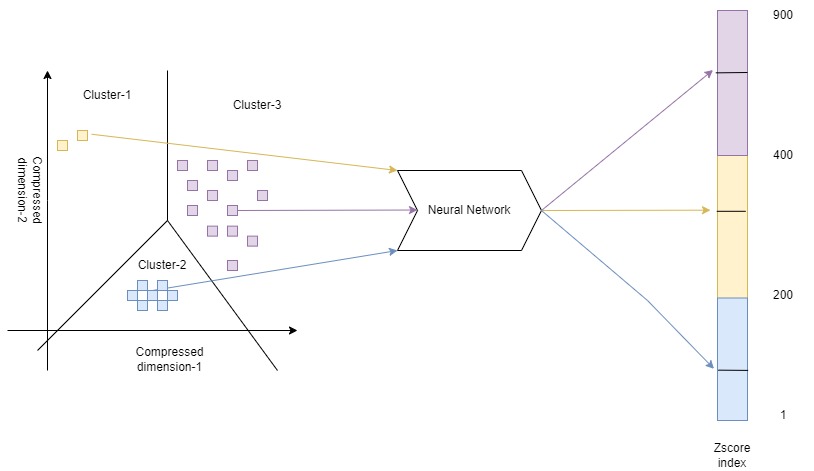} 
	\caption{\textit{\textbf{Overview of the mechanism of assigning users zScores}
		The graph represents users’ onchain history embedded in a 2-dimensional space in which we have been able to separate users into three clusters; each of these clusters has zScore bounds [Section 3.2], and the neural network uses this cluster data along with the user features to predict the user’s zScore [Section 3.3,2]}}
	\label{fig:Model Overview}
\end{figure}

\section{Theory}
This section is divided into four subsections. Section 2.1 explains what zScore represents; Section 2.2 explains clustering and its need; Section 2.3 explains neural networks and the learning paradigm used to train the model; Section 2.4 introduces cryptoeconomic security and Eigenlayer, a protocol we use to achieve cryptoeconomic security. \cite{team2024eigenlayer}.
  
\subsection{zScore}

Before we define zScore, we state definitions of terms and assumptions used. We describe a user to be any wallet that has onchain transactions in any of the verticals of DeFi(Lending Protocols \cite{kaplan2023decentralized}, Decentralized Exchanges(DEXs)\cite{lehar2021decentralized}, Perpetuals \cite{bitstamp2024perp}, Onchain Credit Protocols [\cite{3jane2024introducing}, \cite{etherfi_whitepaper}], or LSTs \cite{gogol2024liquid} and LRTs \cite{carre2024liquid}) captured by our model. For example, a user for our case study on Aave V3 \cite{frangella2022aave} would be any wallet that has opened a position since the protocol's launch. Similarly, we can define users concerning DEXs \cite{lehar2021decentralized} as any wallet that has swapped or provided liquidity since a protocol's launch. 

We define a user's ideal behaviour as any behaviour that does not negatively affect the protocol they have interacted with. Only two actions can negatively affect the protocol, the first being the set of all behaviours for which the protocol must shell out tokens as compensation. A classic example is liquidation calls on lending protocols \cite{lehar2021decentralized}. The second set of behaviours are pumps and dumps and flipper behaviour, which can destroy the reputation of a protocol's token and thereby shed doubt on its validity \cite{clough2023pump}. 

Using the above definitions of users and ideal behaviour, we shall now define onchain reputation. onchain reputation is a quantitative measure of a user's trustworthiness and credibility backed by their historical demonstration of consistent ideal behaviour. This implies that users with perfect historical records of consistent ideal behaviour have good onchain reputations. Similarly, users with consistent demonstrations of non-ideal behaviour would have a bad onchain reputation, meaning low trust and credibility. 

The only assumption we make is that before a user takes any onchain decision, they have access to all the information they need to make an appropriate decision. This basis for the assumption is that all historical price movements, transaction histories, and other required information are publicly available in the blockchain ledger[\cite{nakamoto2008bitcoin},\cite{buterin2014next}].

Using the assumptions and definitions mentioned above, we define zScore as a measure quantifying the user's onchain reputation, which can be any integer between 1-900, with 1 symbolising a consistent display of non-ideal behaviour and 900 symbolising a consistent display of perfect ideal behaviour. Studies like \cite{wolf2022scoring} show user reputation strongly indicates future behaviour. While \cite{wolf2022scoring} restricts itself to a singular protocol, we expand our horizons to all four verticals using our definition of reputation. We argue that a user's onchain reputation strongly indicates their future behaviour. Section 4.4 discusses this point and dives into the utility of zScore, while Section 5 provides a roadmap for protocols to integrate zScore into their reward/incentive systems.

\subsection{Clustering}
Clustering algorithms separate points in an n-dimensional space into different groups. Many different clustering algorithms exist, which differ slightly and are chosen depending on the data's topology and clustering criteria. Classical algorithms like K-NNs \cite{peterson2009k} and Hierarchical Clustering \cite{nielsen2016hierarchical} are commonly used; however, specialised algorithms like Density-Based Spatial Clustering of Applications with Noise (DBSCAN)\cite{khan2014dbscan} and Gaussian Mixture Models (GMMs)\cite{reynolds2009gaussian} are used when the data is dense and improperly separable.

These algorithms fall under the broad category of unsupervised learning algorithms in Machine Learning (ML). Unsupervised Learning\cite{barlow1989unsupervised} refers to the scenario where we don't know the relation between independent variables (inputs) and dependent variables (outputs). Clustering algorithms are used to unearth ties between independent and dependent variables\cite{verma2012comparative} by separating the dataset into groups, which allows us to label each group to a dependent variable(s) or a range of dependent variables, converting unsupervised learning to semi-supervised learning. Algorithms described in [Section 2.3] can be trained only by partially mapping independent and dependent variables \cite{yegnanarayana2009artificial}.

\subsection{Neural Networks}
Artificial Neural Networks \cite{yegnanarayana2009artificial} have been successfully used in various tasks [ImageNet \cite{deng2009imagenet}, NLP \cite{chowdhary2020natural}, Computer Vision \cite{yoo2015deep}]. Our zScoring model also uses a neural network described in [Section 3.3]. The design of our model requires us to use a learning paradigm called Multitask Learning, described in \cite{caruana1997multitask}. In our neural network, we use three components to facilitate multitask learning. The first component is an encoder mechanism described in \cite{wang2016auto}, which helps us extract important, valuable characteristics of the user that are important in determining the zScore. The second component used is an embedding layer, as described in \cite{wolz2010leap}, whose use and role are described in [Section 3.3]. The last component is an attention layer \cite{vaswani2017attention} whose use is also described in [Section 3.3].

\subsection{Cryptoeconomic Security}
\cite{cryptoeconomics_wikipedia} defines cryptoeconomic security as using economic incentives and cryptographic techniques to ensure the security and proper functioning of decentralised networks. Blockchains \cite{nakamoto2008bitcoin} were landmark achievements in the case of decentralised networks. The blockchain Trilemma defines the characteristics of an ideal blockchain, i.e., a blockchain must be secure, decentralised, and scalable. However, research \cite{fu2024quantifying} has shown that most blockchains and other computational models can only satisfy two conditions of the blockchain trilemma at once.

Ethereum \cite{buterin2014next} is a decentralised blockchain network because it establishes cryptoeconomic security through staking \cite{ethereum_pos_documentation}. It has a mechanism called Proof-of-Stake \cite{ethereum_pos_documentation}, which allows validators to pledge ETH tokens as an alibi for trust when they validate blocks, as validators validate more blocks, their stakedETH grows, and so does the trust in the protocol. In Ethereum's Proof-of-Stake system \cite{ethereum_pos_documentation}, if a validator approves a malicious block, a portion of their staked ETH is destroyed as a penalty, a process known as slashing. The slashed ETH is not transferred to other validators but is permanently removed from circulation \cite{buterin2014next}. Slashing is a deterrent to approving malicious transactions, increasing trust in the network. Ethereum \cite{buterin2014next} also introduced smart contracts, which led to the rise of NFTs and other assets \cite{aimultiple_smart_contract_nft}.

The Ethereum Virtual Machine (EVM) allows smart contracts to utilize the cryptoeconomic security of the blockchain. However, integrating other computational models to harness this trust has been challenging. EigenLayer addresses this limitation by enabling protocols to leverage Ethereum's existing validator set and staked capital through a process called restaking. This approach allows new protocols to inherit Ethereum's security without establishing their own validator networks, thereby promoting innovation and expanding the applications of Ethereum's trust network \cite{team2024eigenlayer}.

To achieve the security and decentralization of a complete blockchain like Ethereum \cite{buterin2014next}, protocols with other computational models at their core would often choose to create their own blockchains, which are development and capital-intensive. To combat this problem, Eigenlayer \cite{team2024eigenlayer} introduced the idea of restaking ETH and free-market governance. Restaking ETH allowed validators to stake stakedETH, building off of Ethereum's \cite{buterin2014next} established cryptoeconomic security. Free-market governance allowed protocols with any computational model to integrate with Eigenlayer \cite{team2024eigenlayer} and subsequently become secure by leveraging their vast corpus of validators and Ethereum's \cite{buterin2014next} security. Integrating with Eigenlayer \cite{buterin2014next} is like building a website without any frameworks, i.e., it requires intensive dev-time writing boilerplate code. Hence Othentic \cite{othentic_documentation} has emerged as a plug-n-play framework to allow protocols seamless integration with Eigenlayer \cite{team2024eigenlayer}.

If not already evident, the motivation behind using Eigenlayer \cite{team2024eigenlayer} was to easily construct a cryptoeconomically secure, scalable zScore system without creating our own blockchain. Section 3.4 describes a deep dive into our integration with Eigenlayer \cite{team2024eigenlayer} through Othentic \cite{team2024eigenlayer}.

\section{Methodology}
This section describes our model pipeline and discusses the methodology required to train and implement the model on any protocol. Section 3.1 describes our data-preprocessing strategy, section 3.2 discusses our methods of optimising user clusters and improving validity, and Section 3.3 defines the neural network architecture and the loss functions used to train the network.

\subsection{Data Preprocessing}
Sections 2.1 and 1 highlight the vast domain our system must be able to learn and infer from. As a result, we have multiple sources from which to pull data. We first describe our methodology for selecting data sources, define the structure of the data streams, and then give a general overview of the features we extracted.

We selected protocols which would act as a data source for our model based on their popularity and volume of transactions; for example, the most popular and voluminous lending protocol is Aave V3 \cite{frangella2022aave}, which has been used as a case study in Section 4. Similarly, we have chosen Uniswap V3 \cite{adams2021uniswap} for DEXs; for LSTs and LRTs, we have chosen Lido \cite{lido_documentation} and Eigenlayer \cite{team2024eigenlayer}, respectively, while for onchain credit, we chose ether-fi \cite{etherfi_whitepaper}. 

Most of the datasets we had access to for each of these protocols are publicly available logs fetched from graphQL \cite{quina2023graphql} or the protocols' APIs themselves. These datasets had minimal features needed to describe a transaction between the user and the protocol. As an example, to arrive at a model that calculates zScore, we had to first derive datasets containing features representative of user behaviour rather than transactions, an example is provided in [Section 4.2], and then use clustering to label(assign) users with zScore bounds[Section 3.2]. 

Given that most of our datasets represent a large proportion of users of each data source, we assume they represent the population we are training it for. 

Feature extraction from transaction datasets is done keeping in mind the set of all possible user behaviours. Three different categories of features can easily represent this set of all user behaviours: the first category details the interactions between the user and the protocols in question, the next category is time-specific, i.e., captures the frequencies between each interaction, while the last category captures the volatility of assets used on the protocol. We also have a special category of features which establish interconnectedness between the verticals of DeFi; these features are constructed by following the history of a particular token and seeing how it moves through each vertical. We have multiple hypotheses [Section 6], which we are testing, which have shown a strong interconnectedness between verticals, and capturing such behaviour accurately will allow us to make zScore an accurate representation of user reputation. The case study [Section 4] captures only the first three categories of features; however, it is still an excellent alibi for trust and reputation [Section 4.5]. 

Now that we have a cleaned dataset with the appropriate user-level features extracted, we label each user with their respective zScores using the method described in the next section.

\subsection{Clustering and Partial Labeling}

As described in [Section 3.1], our dataset is unlabelled, and we must cluster users into groups. Since the distribution of users in different verticals differs, the clustering algorithm we use to mine user behaviours might vary. This section describes the general approach we have used in clustering and labelling users; for an example of our vertical-specific approach, we direct the reader to Section 4.2. We first discuss the objective function, which we maximise over to mine user behaviour, before describing our partial labelling framework.

The goal of clustering is to obtain moderately separable clusters, which would help us ensure smooth transitions between zScore ranges when we map "mined" behaviours/reputations to zScores; this approach also allows the neural network model to discover intimate relationships between behaviours [Section 3.3]. Well-separated clusters will set a hard threshold, which might enable unfavourable jumps to creep into the system, making it unpredictable at cluster boundaries. We choose silhouette scoring as an indicator of cluster separability \cite{shahapure2020cluster}. One part of the objective function is the silhouette score \cite{shahapure2020cluster}. Since we aim for moderate cluster separability, we have followed the industry standard, accepting clustering with silhouette scores $> 0.51$ \cite{shahapure2020cluster}. The second component of our objective function is the number of clusters we set an upper bound on during optimisation, explained in the next paragraph. We wish to mine the maximum possible behaviours demonstrated in the dataset while ensuring the appropriate level of cluster separability. Hence, we choose to maximise this objective function. Our objective function is defined mathematically as follows:

\[f(D,l,n) = 10*sil\_score(D,l) + n\]

Where $D$ represents the dataset and $l$ represents the array of cluster labels for each user in the dataset. $sil\_score(D,l)$ represents the function used to calculate the silhouette score \cite{shahapure2020cluster} of the Dataset. $n$ represents the number of clusters. We scale the score by a factor of 10 to ensure we weigh silhouette scores \cite{shahapure2020cluster} higher to prevent poor clustering results. 

We use particle swarm optimisation (PSO) \cite{kennedy1995particle}, which is a genetic algorithm \cite{lambora2019genetic} and has proved effective in such optimisation scenarios \cite{kennedy1995particle}. We optimise over the negative of the objective and control the clustering criteria; for clustering algorithms requiring the number of clusters as an input, we set appropriate lower and upper bounds; the same applies to algorithms which use distance splits and other criteria. Once we find our best-fit clustering, we label users using the framework defined below.

We first label each user with their appropriate cluster and then describe every cluster, using their observed means, standard deviations, ranges and minimum and maximum values of each feature. We use these values to assign each cluster an interval of zScores, which all users in the cluster must fall into. While this process is currently manual, we recognise that this is just an application of domain knowledge and can be automated using other AI models; however, given the system's sensitivity, we choose the manual annotation path for this system iteration. Our inferences are derived using the concerned protocols' whitepapers and expert guidance from those in the domain. Hence, each user is labelled with their respective cluster and the zScore interval they must fall into. In the next paragraph, we set rules on this partial labelling system, which must followed to maintain coherence with our definition of zScore.

Clusters representing users who are "new",i.e. users with few interactions with the protocol, must be ranked around the 100-300 range; this ensures that we are cautious about "new" users while still providing them with the benefit of the doubt. Completely new users, i.e. users with zero interactions/transactions in the respective vertical, must be given a zScore of 100; this is similar to ELO ratings in chess, where players start with a basic rating \cite{berg2020statistical} and progress according to their skill. Similarly, "new" users demonstrating non-ideal behaviour must be penalised heavily and must not be allowed to cross the 200 threshold. Similarly, users that have shown "non-ideal" behaviour at least once must be capped at a zScore of 400 while allocating initial zScores; subsequent ideal behaviour will allow them to progress. zScore ranges for each cluster must be assigned to make it harder to progress to higher zScore ranges, allowing protocols to use it as an alibi for trust. 

The above rules are inspired by classical rating systems such as ELO rating \cite{berg2020statistical}, which have been proven to work over the decades \cite{brusov2021importance}. Our system of assigning zScores allows us to generalise it to any DeFi vertical, provided we know the ideal and non-ideal behaviours. Additionally, this approach will enable us to consider verticals independently before factoring in the interconnectedness of the verticals, allowing our system to be adopted by various DeFi protocols to assess reputation and trust to provide incentives, as mentioned in [Section 5]. 
 
\subsection{Multitask Learning}
Now that the users are partially labelled with their clusters and zScore ranges, we must train a neural network to learn the cluster characteristics and user zScores. This involves using multitask learning \cite{caruana1997multitask} described in [Section 2.3]. The following paragraphs discuss the network architecture used before diving into the loss function we minimise during training. All inputs to the neural network are scaled using the appropriate scaling techniques, consult [Section 4.3] for an example.

We do not discuss the number of neurons or layers in the architecture since they might differ for each DeFi vertical. Instead, we describe the flowchart presented in Figure 2. Our focus is the neural network components mentioned in the model and how user data is transformed to output their respective zScore. The user data is first passed through the shared components, which are then attended by the attention mechanism before being passed on to the respective output components. 

We have two shared components for each of the outputs. The first shared component is a feature extraction auto-encoder whose inputs are only user features, not cluster labels or the respective zScore range. The auto-encoder, as described here \cite{wang2016auto}, enables the neural network to learn the dataset's latent representations, allowing us to capture intricate feature co-dependencies further. The cluster labels are passed to the other shared component, a learnable embedding layer. Each cluster label is mapped to a vector in the layer, and the network learns these embeddings by comparing them to the observed feature importances of the dataset; the loss function for this is described in the following paragraphs. Cluster embeddings allow the neural network to ensure coherence between user features and their cluster labels by mapping the feature importance appropriately. The embedded vector and the extracted features are then passed on to the attention mechanism, which attends to the output appropriately.

Before we move on to discussing the output sections, we provide a mathematical definition for zScore: \[z(f,l) = L[l] \times \sum_{n=1}^ka_nf_n\] where $f$ symbolises the vector of extracted features, $L[]$ symbolises the array of lower bounds of each cluster label $l$. and $a_n$ represents the feature weights and $f_n$ represents the feature values. This formula is constructed to accurately gauge the distance of the user from the lower and upper boundaries of the cluster and assign them a zScore accordingly. The vector $A = <a_1,...,a_n>$ is calculated from the feature weights output head. After training, the weights of the features will correspond to the observed feature importance of each cluster since we embed this information in the embedding layer. The following paragraph describes each of the output heads. 

Using the shared sections of the neural network, we generate two outputs for each user: the user’s score and the feature weights used to derive the zScore. The shared sections are critical when dealing with users near the clusters' boundaries. For instance, if a user lies at the boundary of two clusters, we prioritise attending to the user’s features rather than relying solely on learned cluster features. This approach prevents abrupt changes in zScores between users and ensures a smooth transition across clusters. Conversely, when users are well within a cluster, we can leverage the learned embeddings of that cluster. The score ranges from -1 to 1, while the feature importance output is normalised to fall between 0 and 1. Finally, the score is scaled based on the cluster bounds.

We train the neural network by minimising the loss function of three components: boundary loss, distribution loss, and consistency loss.

\begin{figure}[ht] 
	\centering
	\includegraphics[width=0.7\textwidth]{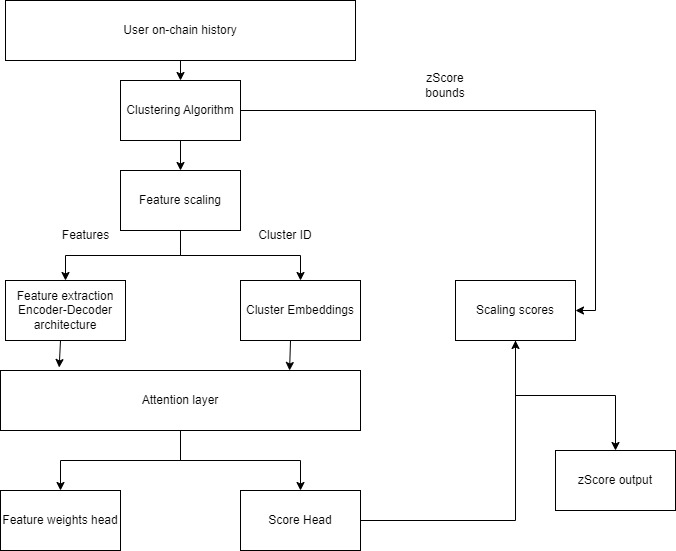} 
	\caption{\textit{\textbf{A flowchart of the model architecture}
		We first transform the user's onchain history into relevant features described in [Section 3.1]; we then classify the user into a cluster [Section 3.2]. The features are then scaled and fed into the neural network [Section 3.3]. The output from the scoring head is then used to scale the score according to the bounds.}}
	\label{fig:Model Flowchart}
\end{figure}

The boundary loss refers to the error reached when the model predicts a zScore outside the acceptable range of the user’s cluster. The distribution loss ensures that the zScores are distributed across the given cluster range and not converging to a singular zScore for all users in that cluster. This loss is made by comparing the ratio of the range of the zScores given by the model to the range of the cluster with our required spread percentage.

The last loss function tests the coherence of the scores concerning the users' positions in the clusters. This is done by comparing the cluster's observed feature importance with the model's feature importance output. The coherence test ensures that the model is learning the cluster representations in its embedding layer and also learning to pay attention to the right features for the outputs.

\subsection{Establishing Crypto-Economic Security}

As mentioned in [Section 2.4], Eigenlayer \cite{team2024eigenlayer} is a protocol allowing us to build off of established cryptoeconomic security while being scalable. We choose to use EigenLayer to deploy our Autonomous Verifiable Service
(AVS) \cite{eigenlayer_avs_developer_guide}. This section describes the mechanism behind our execution service and Eigenlayer \cite{team2024eigenlayer} integration through Othentic Registry \cite{othentic_documentation}. A visual representation of our decentralized system is provided in Figure 3. The following paragraphs describes the workflow. 

The zScore AVS continuously gathers transaction logs from Protocols (Lending protocols \cite{lehar2021decentralized}, DEXs \cite{kaplan2023decentralized}, LSTs \cite{team2024eigenlayer}) across multiple blockchain networks. This process occurs in intervals depending on the frequency of updating the zScores, during which user transactions are extracted and stored in a database. The collected wallet data is input for the execution service, which computes and validates user zScores. In the execution service, once we have the transaction logs for users, we first transform these logs into features accepted by the AI model described in the previous section. 

The conventional route to publishing these computed zScores onchain would have been through smart contracts \cite{buterin2014next}. However, a major roadblock in this method prevented us from doing so. By publishing user zScores as smart contracts, we would take the responsibility of covering the gas costs for the users, which would quickly bleed our coffers dry since we estimated our total gas cost for 1.3 million Aave v3 \cite{frangella2022aave} to be $US\$10,000$ per update. To solve this problem, we devised a workaround involving Merkle Tree \cite{kuznetsov2024merkle} and Databases \cite{rocksdb}, significantly reducing costs while maintaining the same level of security.

Computed zScores of users are stored in a database called RocksDB \cite{rocksdb}, which has fast and efficient storage. These features make RocksDB a suitable foundation for building a Merkle-Tree \cite{kuznetsov2024merkle} based database. 

Once all zScores are computed, we then obtain the Merkle-root \cite{kuznetsov2024merkle} and pass it on to the validators for proof-of-task.  Validators validate proof of task (Merkle root \cite{kuznetsov2024merkle}) by random sampling the users and try to recompute the Merkle root \cite{kuznetsov2024merkle} using the stored user zScores and their Merkle hashes \cite{kuznetsov2024merkle}. If the computation is not malicious, i.e. at least a two-thirds quorum is reached on its validity, we publish the root with updated zScores to the smart contract, which is deployed to multiple blockchains. This process is repeated periodically. 

The Othentic registry contract \cite{othentic_documentation} acts as a middleware for interacting with EigenLayer \cite{team2024eigenlayer}. It facilitates secure verification of zScore computations by providing a standardized interface for task validation, ensuring that operators within the EigenLayer \cite{team2024eigenlayer} framework can authenticate and attest to the integrity of credit scores
.
By leveraging EigenLayer’s \cite{team2024eigenlayer} decentralised security model, zScore AVS ensures that credit scores are computed, validated, and propagated across chains in a transparent, verifiable, and tamper-resistant manner.Additionally, anyone can verify these scores by requesting the AVS \cite{eigenlayer_avs_developer_guide} for the score through an API with Merkle hashes \cite{kuznetsov2024merkle} and verifying that the Merkle root \cite{kuznetsov2024merkle} is the same Merkle root \cite{kuznetsov2024merkle} onchain.

\begin{figure}[ht]
	\centering
	\includegraphics[width=0.9\textwidth]{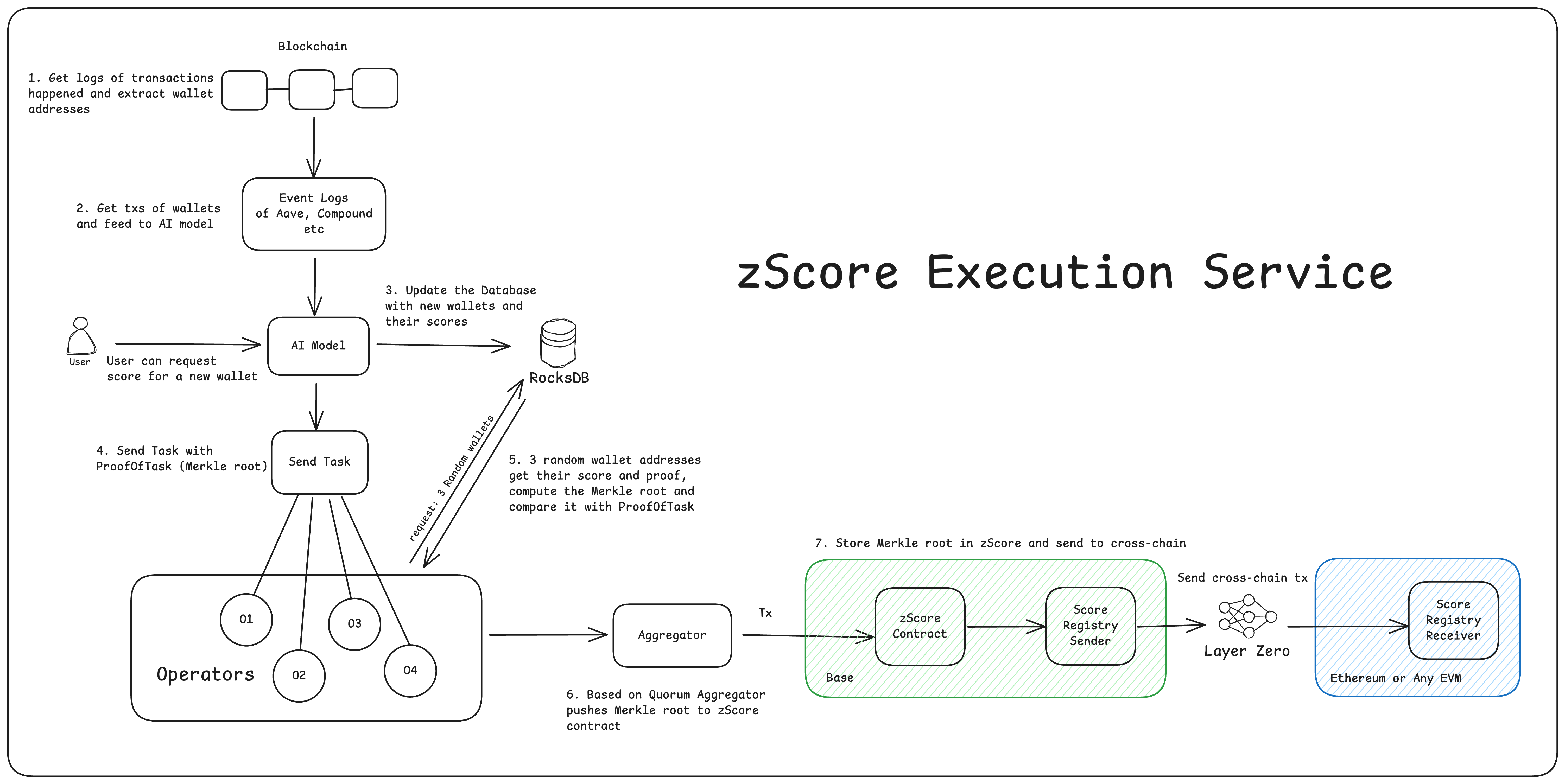} 
	\caption{\textit{\textbf{Flowchart of zScore Execution service}
	    zScore AVS, we first fetch user logs and then extract features from it before passing them to the model [Section 3.3]. We then store user features and their zScores in a DB, which we then generate the Merkle root for. The Merkle root is then validated by the validators, and once quorum is reached, we publish it to Base [Refernce].
        }}
	\label{fig:Model Overview}
\end{figure}

\section{Case Study: Aave V3 Accounts}
This section implements our zScore model on users on Aave V3 \cite{frangella2022aave}, implemented and available live at [myzscore.ai]. The first section formally defines the problem of assigning users their zScores by stating assumptions and defining key terms that will help define non-ideal behaviour in terms of lending protocols in general. We then extract relevant features as described in [Section 3.1] in [Section 4.2] before clustering users and training the neural network model in [Section 4.3]. Subsequently, we provide a comparative study between our scoring system and other protocols in [Section 4.4] before concluding with a real-world scenario where our model would have saved Aave V3 \cite{frangella2022aave} from considerable losses in [Section 4.5].

\subsection{Problem Setup}
This subsection is divided into three parts. First, we define a user for lending protocols. Then, we define terms and state assumptions that will allow us to describe ideal behaviour as the negation of non-ideal behaviour accurately. Once we have defined perfect behaviour, we describe our raw dataset and use [Section 3.1] to convert it into a dataset representing behaviour.

In the context of lending protocols \cite{lehar2021decentralized}, a user is any wallet that has made any of the wallet-level calls, i.e., Borrow, Repay, UsageAsCollateral, Deposit, or LiquidationCall \cite{frangella2022aave}, at any point in its history. We isolate transactions between the wallet and any lending protocol and consider this a user. To accurately define ideal behaviour, we define its complement, i.e., non-ideal behaviour. To do so, we shall first define positions in lending protocols, the volatility of coins in lending protocols, and a user's health factor \cite{frangella2022aave}. 

A "position" is any borrowing or repayment activity for the same coin; positions are closed when the debt reaches zero. Repayments reduce the debt, while borrowing adds to it. 

Aave V3 \cite{frangella2022aave} defines the Liquidation Threshold as the point at which a user's collateral becomes uncollateralised and subject to Liquidation. Higher Liquidation Thresholds imply lower trust in the coin's stability \cite{lehar2021decentralized}. Hence, we can use the liquidation thresholds to measure the coin's volatility. We first obtain the Liquidation Thresholds of coins across all blockchains$^1$; we compute the average if a coin has multiple Liquidation Thresholds. We then arrange the coins in ascending order based on their Liquidation Thresholds; the top fifty percentile are considered non-volatile coins, and the bottom fifty percentile are considered volatile coins. Using Liquidation Thresholds to decide coin volatility allows us to differentiate between non-volatile and volatile coins since non-volatile coins have lower thresholds than volatile coins. 

Aave v3 \cite{frangella2022aave} and other lending protocols \cite{frangella2022aave} are structured to incentivise users who cover the Liquidated Amount when a LiquidationCall is made. To judge if a user may be liquidated, they use a formula called Health Factor, which is described below: \[Hf(w) = \frac{\text{Total Collateral Value} \times \text{Weighted Average Liquidation Threshold}}{\text{ Total Borrow Value}}\] where $w$ represents the user. The $Hf(w)$ going below $1$ indicates that our Total Borrow Value exceeds the accepted limit, which implies that the user may be subject to liquidation. A user at risk of Liquidation suggests that the protocol will have to shell out tokens when they are Liquidated \cite{frangella2022aave}. This behaviour is non-ideal according to our definition in [Section 2.1]. 

We define ideal behaviour as any behaviour that prevents the possibility of the health factor falling below one and non-ideal behaviour as any behaviour that increases the possibility of the health factor falling below 1. Since reputation is a consistent display of ideal behaviour, and a good reputation implies a higher zScore, users with high zScores should have 0 or negligible instances of their Health Factor falling below 1.  

Our data source was the transaction data of all Aave V3 \cite{frangella2022aave} users across all blockchains retrieved using graphQL \cite{quina2023graphql}. The transaction data had a separate file for every blockchain for every possible call (Borrow, Repay, Deposit, UsageAsCollateral, LiquidationCall). The data was represented by a few features: BlockId, Wallet\_Address, Amount, Call, Timestamp, and Coin. Using these features and our methodology mentioned in [Section 3.1], we converted them to the three categories of features mentioned in [Section 3.1]. A simple analysis of the feature-engineered dataset, referred to as dataset from now, showed that we had approximately $100,000$ users, out of which only $3,333$ had undergone a liquidationCall and about $46,000$ were relatively new users, i.e., users with $<10$ Borrow calls. The characteristics of these types of users are described in the next section. We dropped 10 users who had negative values in their time-specific features. We then split the dataset into two; one contained non-LiquidationCalls, while the other contained users who had at least one LiquidationCall. We then clustered users in both these datasets, as described in the next section.

\subsection{Implementing Clustering}
Following the methodology we described in Section 3.2, we select three clustering algorithms, namely K-Means \cite{peterson2009k}, Agglomerative Clustering \cite{nielsen2016hierarchical} and DBSCAN \cite{khan2014dbscan}, to cluster users in both datasets. We set the lower bound in the number of clusters to be 10 and an upper bound to be 50 for the dataset containing users with zero LiquidationCalls(referred to as non-liquidation set) and a lower bound and upper bound of 5 and 20, respectively, for the dataset containing users with non-zero LiquidationCalls (referred to as liquidation set). We set the lower bounds to ensure we have at least 10 progression intervals for zScore. However, this is not a hard threshold; instead, we picked the number of clusters which maximised the objective function described in Section 3.2. 

We applied the PSO algorithm \cite{kennedy1995particle} with 30 particles for the non-liquidation set to enable faster convergence in the number of clusters and 10 particles for the liquidation set. The PSO algorithm \cite{kennedy1995particle} showed that K-Means \cite{peterson2009k} was the most optimal method for both datasets. DBSCAN \cite{khan2014dbscan} consistently failed to classify a significant portion of users and Agglomerative clustering \cite{nielsen2016hierarchical} yielded sub-optimal results. The clustering we obtained for the liquidation set had a silhouette score \cite{shahapure2020cluster} of $0.59$ with a total of $10$ clusters. The clustering we obtained for the non-liquidation set had a silhouette score \cite{shahapure2020cluster} of $0.60$ with $23$ clusters. However, we had $>76\%$ of users concentrated in a cluster with significantly larger ranges of values observed in features. We decided to split this cluster into sub-clusters; using the same methodology, we obtained a sub-clustering with a silhouette score \cite{shahapure2020cluster} of 0.59 with 19 sub-clusters. The distribution of these clusters is shown in Figure 4. 

\begin{figure}[ht] 
	\centering
	\includegraphics[width=0.6\textwidth]{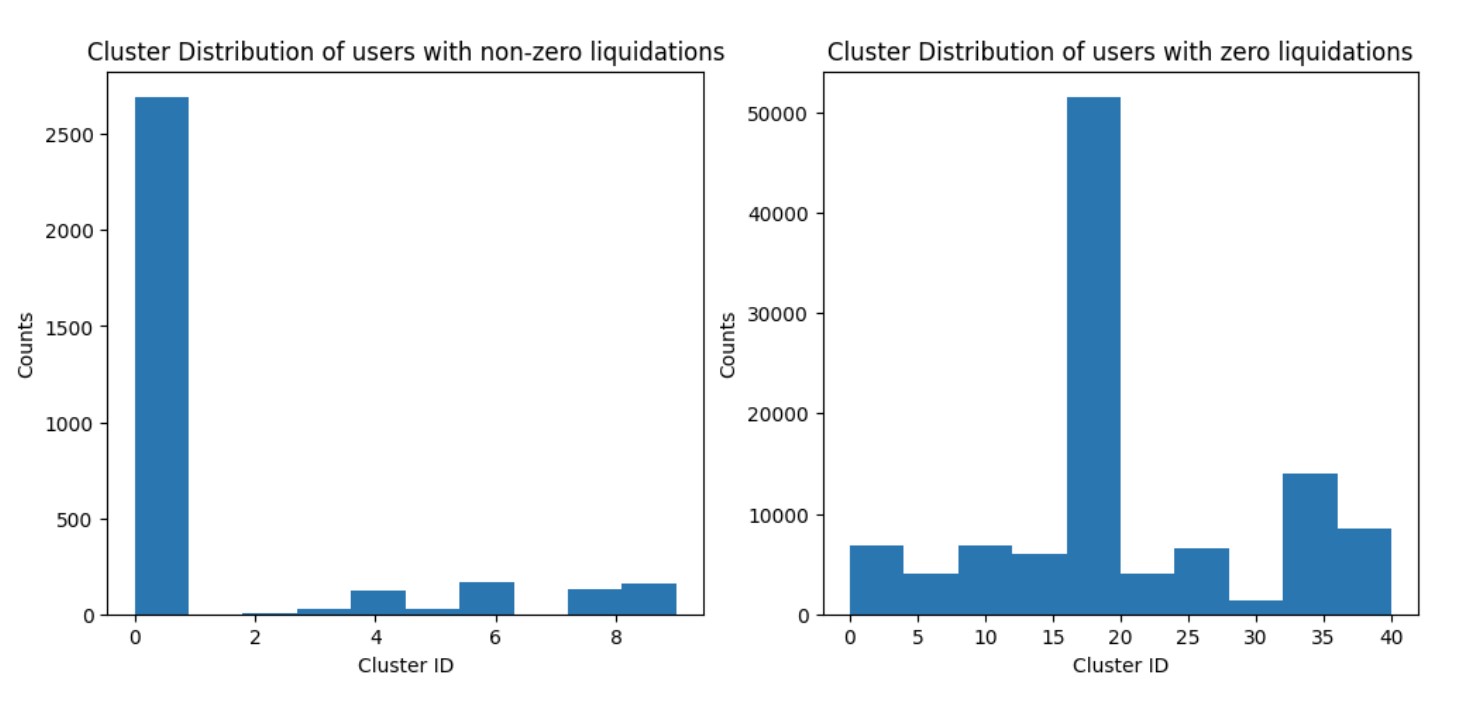} 
	\caption{\textit{\textbf{Distribution of users across clusters}
    Left - the distribution of users with non-zero liquidations. Right - the distribution of users with zero liquidations. We have assigned the subclusters of cluster0 IDs from 0-18 [Section 4.2]}}
	\label{fig:Model Overview} 
\end{figure}

According to our methodology, the next step was to label users with their cluster-specific zScore interval. Following the system described in [Section 3.3], we label clusters representing new users with zScore ranges of 100-250; we also cap the maximum possible zScore of the liquidation set to 400. We then allow zScore ranges to new users who have undergone LiquidationCalls. These ranges had a maximum upper bound of $<150$. All the other range allotment which we did was done using the following importance list: We prioritised the user's frequency of interactions and Call counts equally as this revealed how consistent they were with their interactions; we looked at their LiquidationCalls and proportion of volatile investments both in Deposits and Borrows. Users with sporadic interactions were given lower zScores, as were users with many transactions involving volatile coins. The following section describes training the neural network before moving on to zScore analysis.

\subsection{Training the Neural Network}
We implemented the neural network architecture and set up the loss function, which we minimise during training, according to [Section 3.3]. We use early stopping \cite{ji2021early} to prevent overfit in the models and train two separate models for the two datasets. We followed a 70-30 training validation split for both models. For each model, we chose the batch size by iterating over the following batch sizes: $\{64,128,256,512,1024,2048,4096,8192\}$, the most optimal batch size was the one which yielded the minimum validation loss. We found that batch sizes of $256, 4096$ were the best for the liquidation and non-liquidation sets, respectively. For both models, we used a patience \cite{prechelt2002early} counter of 15. 

In both sets, our zScore interval for each cluster had a maximum range of $100$ points. Hence, keeping our zScore within bounds was essential to avoid an error of $>100$ when classifying users. To do so, we weighted bound loss the most and weighted coherence and distribution loss equally in most cases. For clusters with few users, we chose to skip data augmentation [Section 6], with somewhat weighted distribution loss and bound loss being the highest, followed by coherence. The basis for this is that distribution loss would force the embedding layer \cite{wolz2010leap} to learn the cluster embedding properly while the bound loss would ensure prediction within bounds; given the small number of users in these clusters, we would not have to worry about coherence as long as users are located around the centre, i.e. the cluster is compact. Similarly, we weighted all three components equally in clusters, with many users concentrated at the centre and others lying closer to boundaries. 

We first scaled the user features appropriately, i.e. counts were scaled using $logexp$ while time and other continuous features were scaled using normalisation before we began training. During training, we observed that the bound loss converged to $0.0$ for both validation and training subsets, while the coherence and distribution losses converged to significantly low values before early stopping. We noticed that the main reason we could not get coherence and distribution losses closer to $0.0$ was an imbalance in some clusters; this has been improved in the next iteration [Section 6]. 

\subsection{zScore Analysis}
This section first describes the distribution of zScores for the $103,000$ users from Aave V3 \cite{frangella2022aave} before we compare our model with Cred Protocol \cite{wolf2022scoring}, which has done related work only in lending protocols. We discuss inferences we drew from their work, differences, and advantages we have over their implementation. 

Figures 5 \& 6 show the distribution of zScores for all possible clusters. We observe convergence to a singular zScore or a small interval of scores in clusters with fewer than 10 users. In large clusters with low feature variances, we observed convergence to zScores, too. However, most clusters had uniformity in their distributions and the required spread. The overall distribution of $~103,000$ users we used to train the network is shown in Figure 7. We see that most users are between the 50-200 range, which implies they are relatively new users or must have undergone a LiquidationCall in their history. Additionally, we see that $<20\%$ of users were able to cross the $600$ point threshold, indicating that our scoring system behaves just like its theoretical definition in [Section 2.1]. We can also conclude that zScore is an accurate mapping of the distribution of users in Aave v3 \cite{frangella2022aave} and not a transformed mapping, which may introduce biases towards a particular cluster of users; this is important when we compare ourselves with Cred \cite{wolf2022scoring} in the following paragraphs.

\begin{figure}[ht]
	\centering
	\includegraphics[width=0.8\textwidth]{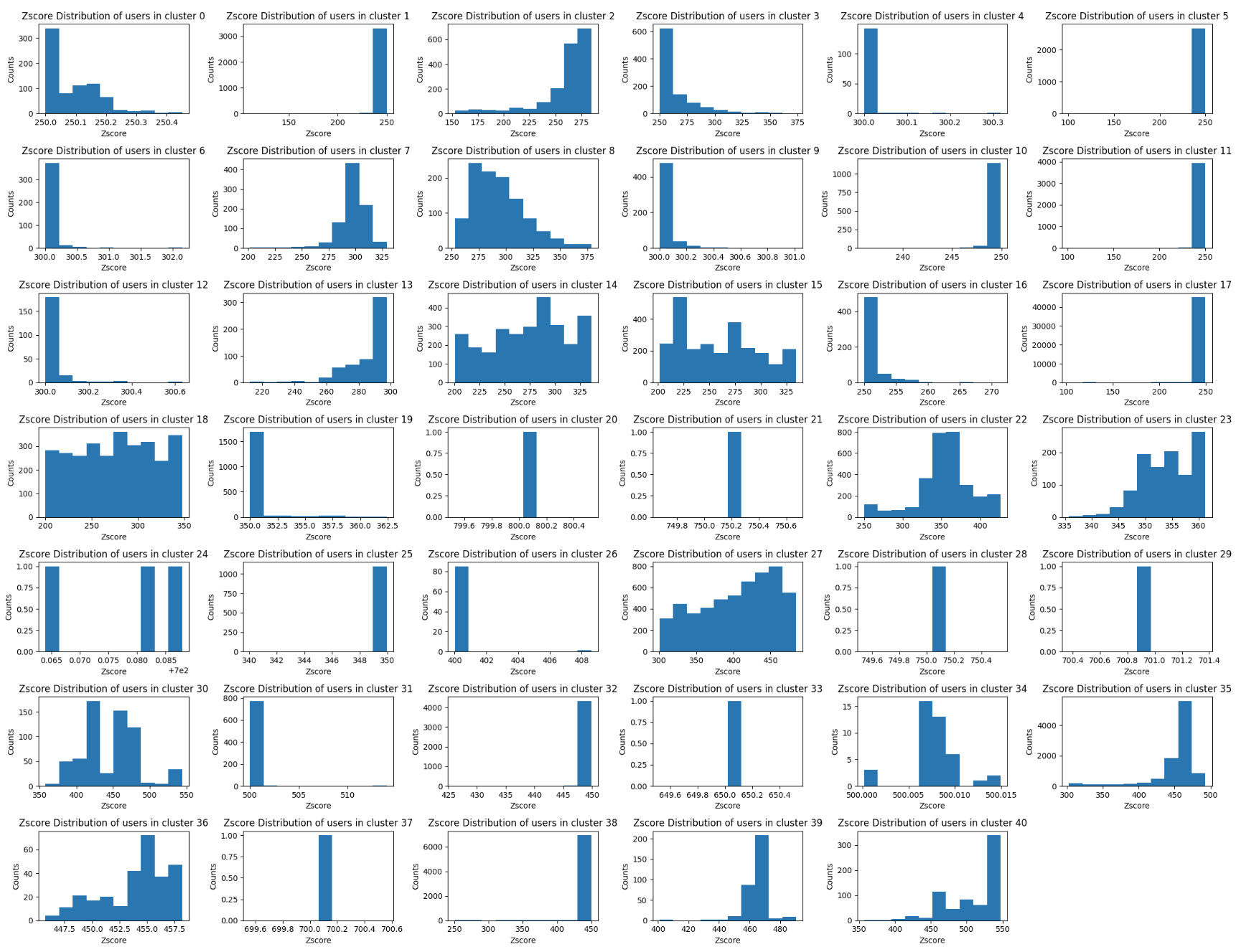} 
	\caption{\textit{\textbf{Cluster-wise distribution of users with non-zero liquidations [Section 4.4]}
    Almost all clusters have skewed distributions, with clusters having few users converging to a small range of zScores. An important insight, however, is the convergence of zScores of new users to a range between 100-250. 
    }}
	\label{fig:Model Overview}
\end{figure}

\begin{figure}[ht]
	\centering
	\includegraphics[width=0.7\textwidth]{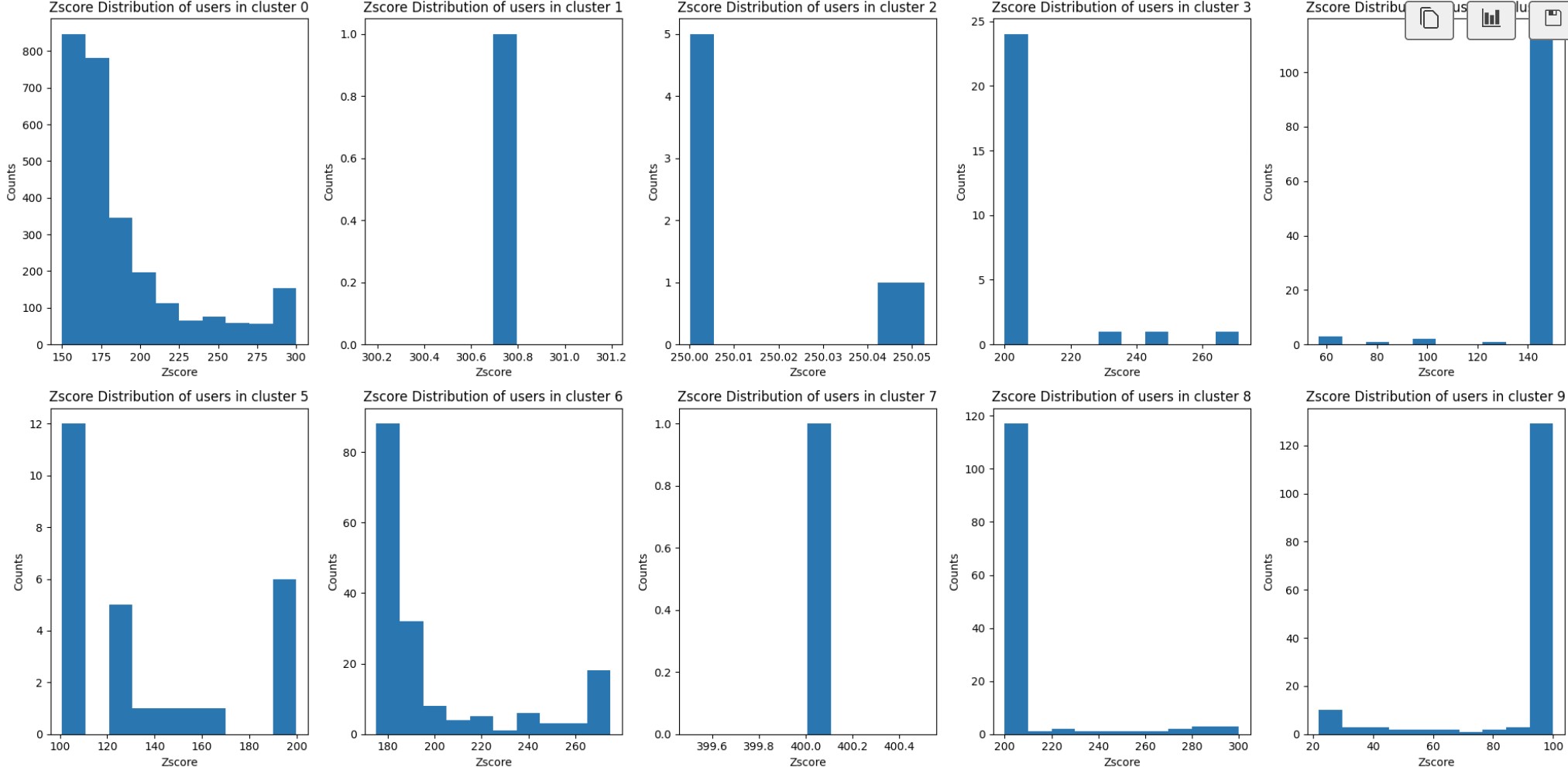}
	\caption{\textit{\textbf{Cluster-wise distributions of users with zero liquidations [Section 4.4]}
		All clusters have skewed distributions, with most clusters having a large proportion of users towards the endpoints of their zScore bounds.}}
	\label{fig:Model Overview}
\end{figure}

\begin{figure}[ht]
	\centering
	\includegraphics[width=0.6\textwidth]{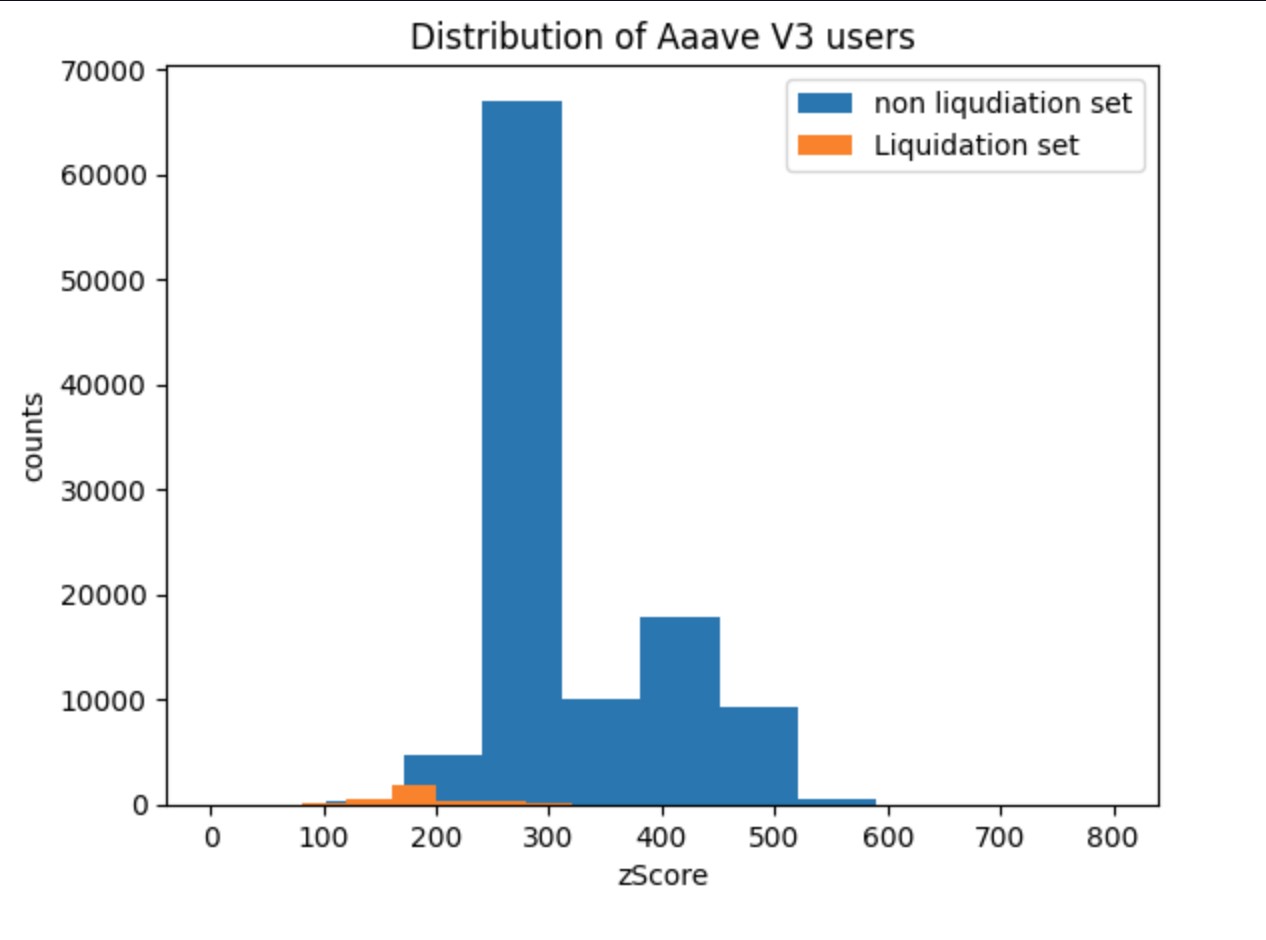}
	\caption{\textit{\textbf{Distribution of all Aave V3 Users used to train our model}}}
	\label{fig:Model Overview} 
\end{figure}

Cred Protocol \cite{wolf2022scoring} was one of the first companies to attempt credit scoring users on Aave V2 \cite{}. We use multiple inferences from Cred \cite{wolf2022scoring} while constructing the scoring model for Aave V3 \cite{frangella2022aave}. The first inference we use is that past user behaviour/reputation is a strong indicator of future behaviour/reputation. This justifies our use of only using onChain history to calculate zSCore. Lending protocol-specific inferences: LiquidationCalls and HealthFactor are an indicator of risk and increase the chance of LiquidationCalls on a position, allowing us to define ideal and non-ideal behaviour with respect to reputation in [Section 4.1].

However, Cred \cite{wolf2022scoring} have its own set of drawbacks, which our model eradicates, allowing our model to improve theoretically and practically significantly [Section 4.5]. Cred \cite{wolf2022scoring} calculates the probability of a new position causing the $Hf$ of the user to fall below one in a $90$ day window. Their model does not consider that the user might take evasive action to ensure the $Hf \ge 1$. Overlooking this crucial fact would imply inaccurate probabilities of Delinquencies, as defined by Cred \cite{wolf2022scoring}. Our model, however, has no such constraint, and its score accurately represents reputation and credibility \cite{sahoo2025tweet}. 

Cred \cite{wolf2022scoring} trains their model only on closed positions and does not include open positions, reducing the number of users significantly. Additionally, they fail to mention how they handle new users and what score ranges they would ideally fall into. A deeper analysis of their formula to calculate probability reveals that new users will most certainly be ranked higher than older users with fewer Liquidations. This hypothesis is further strengthened when we look at the FICO score distribution, to which the probabilities are mapped. We know the distribution of users on Aave V2 \cite{aave_v2_whitepaper} and mapping that to FICO scores \cite{arya2013anatomy}, we see that most of it is concentrated in the higher ranges; looking at their probability vs score chart \cite{wolf2022scoring}, we notice that new users with no liquidations must have lower risk and hence be ranked higher. The fundamental problem is that this allows malicious actors to exploit the low-risk profile and default. zScore, on the other hand, does not place new users on such a high pedestal; instead, we allow them to start from a baseline and earn their reputation through demonstrations of idea behaviour, making it more practical and trustworthy \cite{sahoo2025tweet}.

Cred's model is a proprietary ML model \cite{wolf2022scoring}. This methodology lacks transparency due to the opacity of proprietary models Furthermore, Cred Protocol's credit scoring is delivered through centralized infrastructures, which are not verified by any third-party verification (like a network of peers which verify the execution and storage of data with a cryptoeconomic security model) potentially leading to single points of failure and biases in credit assessments and distribution. In contrast, zScore is deployed as an AVS \cite{eigenlayer_avs_developer_guide} on top of EigenLayer, utilising its economic security \cite{cryptoeconomics_wikipedia} to validate its credit scoring predictions and utilizing operators to verify the correctness of its AI model. This process involves using Merkle proofs \cite{kuznetsov2024merkle}, enabling operators to confirm that the infrastructure distributes correct scores. By leveraging EigenLayer's \cite{team2024eigenlayer} cryptoeconomic security \cite{cryptoeconomics_wikipedia} and network of operator nodes, zScore ensure that its credit assessments and distribution are both transparent and decentralised.

\subsection{zScore: An Alibi for Creditworthiness}
In this section, we present our model's utility in efficiently capturing users' reputation on the onchain economy. On February 3rd, 2024, markets experienced a bloodbath, which led to the biggest liquidation event on Aave V3 \cite{frangella2022aave}; this was caused by several factors; however, since this was not a black-swan event, not all users were affected. We analysed over $5,000$ wallets which were liquidated and noticed that more than $75\%$ of liquidations had occurred in users with zScores of $<300$. There were about $3\%$ of liquidations from users between $400-600$ ranges indicating some risk involved. However, users with lower zScores around $ 10-20 $ also underwent liquidations. Section 4.1 defines LiquidationCalls as non-ideal behaviour and a threat to a user's reputation. Additionally, our scoring system, defined in [Section 2.1] highlights the high risk factor of new-users and users in zScore intervals $<300$, proving that our system has some credibility. The liquidations in the $400-600$ range were negligible; however, since these were zScores before users got liquidated, most of the users would have seen a drop of at least $200$ points since users with non-zero liquidations are capped at a zScore of $400$ [Section 4.2]. A detailed analysis is presented in \cite{sahoo2025tweet}, a post by one of the co-authors. 

This event \cite{sahoo2025tweet} was a litmus test for our model, and it was able to accuratly pinpoint users with tendencies to display non-ideal behaviour. We now discuss how we could utilize zScore to cut protocol losses and allow incentives to users with good reputations in the next section. This reputation based integration will open new doors for the onchain economy by increasing capital efficiency, 

\section{Applications}
This section explores practical applications for the zScore system across the crypto ecosystem. A decentralized reputation system can be massively useful in incentive distribution, onboarding real users, rewarding loyal and value adding users and much more. We demonstrate how our scoring model can enhance existing DeFi infrastructure while enabling novel functionalities previously unfeasible due to the lack of reliable onchain reputation assessment.

\subsection{Lending Protocols}
The most immediate application of zScore lies in lending protocols \cite{lehar2021decentralized} which can significantly enhance risk assessment and capital efficiency. Our system enables three key innovations in lending, which are discussed below.
\begin{enumerate}
    \item Dynamic Loan-to-Value (LTV) Ratios : Protocols can implement variable LTV ratios \cite{frangella2022aave} based on user zScores, allowing higher leverage for users with demonstrated repayment history. This creates a more efficient capital market while maintaining system security through data-driven risk assessment.
    \begin{enumerate}
        \item Example: Aave v3 \cite{frangella2022aave} currently employs fixed LTV ratios to determine borrowing limits based on collateral. By incorporating zScore, Aave v3 \cite{frangella2022aave} could offer dynamic LTV ratios, allowing users with higher zScores—indicative of strong repayment histories—to access higher LTVs. This adjustment would enable such users to borrow more against their collateral, enhancing capital efficiency while maintaining protocol security.
    \end{enumerate}
    \item Interest Rate Optimization: By incorporating zScore into interest rate models, protocols can offer preferential rates to users with strong credit/loan repayment histories. This approach mirrors traditional finance practices without predatory practices.
    \begin{enumerate}
        \item Example: Morpho \cite{delaunay2023morpho} is a peer-to-peer layer built on top of lending protocols like Aave V3 \cite{frangella2022aave} and Compound \cite{leshner2019compound}, aiming to improve rates for both lenders and borrowers. Integrating zScore into Morpho's \cite{delaunay2023morpho} system could allow for personalised interest rates, offering lower rates to users with higher zScores. This approach would reward responsible borrowers and enhance the protocol's competitiveness.
    \end{enumerate}
    \item Under-Collateralized Lending: For users with exceptional zScores (e.g. $>700$), protocols could offer under-collateralized loans, marking a significant step toward true DeFi \cite{jensen2021introduction} credit markets. This should be implemented gradually, with careful monitoring of protocol health metrics as zScore is introduced, higher LTVs \cite{kaplan2023decentralized} are offered, variable interest rates are offered, and the effects are seen over time.
    \begin{enumerate}
        \item Example: 3Jane \cite{3jane2024introducing} is pioneering credit-based money markets, enabling users to borrow against their creditworthiness and future yield without full collateralisation. Using zScores, 3Jane \cite{3jane2024introducing} can assess borrowers' credit risk more accurately, facilitating under-collateralized loans for users with exceptional zScores (e.g., above 700). This strategy promotes capital efficiency and broadens access to credit within the DeFi \cite{jensen2021introduction} ecosystem.
    \end{enumerate}
\end{enumerate} 

\subsection{DEXs (Decentralized Exchanges)}
Implementing behavioural scoring in DEXs \cite{lehar2021decentralized} enables precise fee optimisation and enhanced trading dynamics. Below are three core innovations this system enables:

\begin{enumerate}
    \item Dynamic Fee Tiers: Exchanges can implement variable fee structures calibrating user trading patterns. Long-term holders with high behavioral scores (e.g., consistent trading volumes, minimal sandwich attacks) receive reduced fees, potentially as low as 0.1\% versus standard 0.3\%. This rewards sustainable trading practices while maintaining protocol revenue. 
    \begin{enumerate}
        \item Example: Uniswap v4 \cite{adams2021uniswap} introduces customizable hooks, enabling the implementation of dynamic fee structures. This flexibility allows for variable fee tiers based on user behavior. For instance, long-term holders with high zScores could benefit from reduced swap fees, such as 0.1\% compared to the standard 0.3\%, incentivizing sustainable trading practices while maintaining protocol revenue.
    \end{enumerate}
    \item Maximal Extractable Value (MEV) \cite{cowdao_mev_protection} Protection Prioritization: Users with established positive trading histories gain priority access to MEV protection features \cite{cowdao_mev_protection}. This includes preferential routing through aggregators and enhanced slippage protection, creating a more equitable trading environment based on demonstrated behavior rather than just transaction size.
    \begin{enumerate}
        \item Example: CoW Swap \cite{cow_swap} offers native MEV protection by utilizing batch auctions and off-chain order matching, reducing the likelihood of front-running and sandwich attacks. By integrating zScores, CoW Swap \cite{cow_swap} could prioritize users with positive trading histories, granting them enhanced access to MEV protection features \cite{cowdao_mev_protection}.
    \end{enumerate}
    \item Liquidity Provider Incentives: The system enables targeted LP rewards based on liquidity provision history. Providers demonstrating stable, long-term liquidity commitment receive enhanced yield incentives, improving pool stability and reducing impermanent loss risks across the protocol.  
    \begin{enumerate}
        \item Example: Curve Finance \cite{curve_dex_pools} employs a reward system that benefits liquidity providers (LPs) \cite{lehar2021decentralized} based on their contribution and duration of liquidity provision. By incorporating zScores, Curve \cite{curve_dex_pools} could further refine its incentive mechanisms, offering enhanced yield incentives to LPs \cite{lehar2021decentralized} with stable, long-term commitments. This strategy would contribute to deep liquidity.
    \end{enumerate}
\end{enumerate}

\section{Conclusion}
We have presented and demonstrated a novel method of quantifying user reputation on onchain economies. We showcase the utility of zScore [Section 4.5] and describe possible applications which would improve existing systems in Decentralized Finance [Section 5]. zScore, however, is still evolving, and we have large-scale improvements that fall into two broad categories. Our first category is data-centric, where our main goal is to be able to capture inter-vertical user behaviour and examine its effects on zScores. We hypothesize that a user's reputation is universal, hence, they must display similar behaviour across domains. We have started work on validating our hypothesis, and our next article will describe the results. Another improvement in the same category is to be able to design a feature-engineering pipeline flexible enough to adapt to new protocols that emerge in a vertical. Data augmentation in the case of imbalanced clusters is also an improvement we have made, we are ready to start testing and comparing model performances. Integrating all verticals and establishing interconnectedness will lead to an explosion of features that will make it harder to cluster and "mine" behaviours, hence, we must find an algorithm that is stable and converges at high dimensions. 

The other category of improvements caters to the implementation of zSCore, in this category, we propose a continuous scoring model built on our model, the idea is that given a user's score and a set of new transactions, we must be able to assign a new zScore to the user without calculating their whole history.

\clearpage

\bibliography{science_template} 
\bibliographystyle{acm}

\section{Acknowledgements}
We would like to acknowledge Devon Martens, Derek Silva, and the rest of the zeru team for proofreading the draft of the paper and providing invaluable feedback. We thank Derek Silva for his eye for detail in finding errors. 

\end{document}